\documentclass[conference]{IEEEtran}
\IEEEoverridecommandlockouts
\usepackage{amsmath,amsfonts}
\usepackage{algorithmic}
\usepackage{algorithm}
\usepackage{array}
\usepackage[caption=false,font=footnotesize,labelfont=rm,textfont=rm]{subfig}
\usepackage{textcomp}
\usepackage{stfloats}
\usepackage{url}
\usepackage{verbatim}
\usepackage{graphicx}
\usepackage{cite}
\usepackage{makecell}
\usepackage{booktabs}

\makeatletter

\newcommand{\Rmnum}[1]{\expandafter\@slowromancap\romannumeral #1@}
\makeatother

\def\BibTeX{{\rm B\kern-.05em{\sc i\kern-.025em b}\kern-.08em
    T\kern-.1667em\lower.7ex\hbox{E}\kern-.125emX},maxnames=4}
\begin{document}

\title{Full-Angle Ray Antenna Array and Omnicell Wireless Communication System\\
}
\author{
	\IEEEauthorblockN{
		Xuancheng Zhu\IEEEauthorrefmark{1}, 
		Zhiwen Zhou\IEEEauthorrefmark{2}, 
		Yong Zeng\IEEEauthorrefmark{2}\IEEEauthorrefmark{3},} 
	\IEEEauthorblockA{\IEEEauthorrefmark{1}School of Information Science and Engineering, Southeast University, Nanjing 210096, China}
	\IEEEauthorblockA{\IEEEauthorrefmark{2}National Mobile Communications Research Laboratory, Southeast University, Nanjing 210096, China}
	\IEEEauthorblockA{\IEEEauthorrefmark{3}Purple Mountain Laboratories, Nanjing 211111, China} 
	Email:\{213223563, zhiwen\_zhou, yong\_zeng\}@seu.edu.cn
} 	

\maketitle

\begin{abstract}
	Ray antenna array (RAA) was recently proposed as a novel multi-antenna architecture that arranges multiple massive cheap antenna elements into  \emph{simple uniform linear arrays} (sULAs) with different orientations. Compared with traditional architectures like hybrid analog/digital beamforming with \emph{uniform linear array} (ULA) and \emph{uniform circular array} (UCA), RAA has several promising advantages such as significantly reduced hardware cost, higher beamforming gains and the ability of providing uniform angular resolution for all directions.
	In this paper, we propose a full-angle RAA architecture and an innovative omnicell wireless communication paradigm enabled by full-angle RAA. 
	The proposed full-angle RAA expands RAA's orientation angle to the full angle domain, such that the RAA's advantages can be exploited to all directions. 
	This further enables the new concept of omnicell wireless communication system, with the base station equipped by full-angle RAA and deployed at the center of each cell. Compared to the conventional cell sectoring wireless communication system, the proposed omnicell system is expected to not only significantly reduce the inter-user interference, but also improve the cost efficiency.
	Extensive analytical and numerical results are provided to compare those key performance indicators such as the spatial resolution and the communication rate of the proposed full-angle RAA based omnicell wireless communication system against the conventional ULA/UCA-based cell sectoring systems. 
\end{abstract}

\section{Introduction}
	Multi-antenna or multiple-input multiple-output (MIMO) technology acts as a milestone in the evolution of wireless communication systems. Earlier MIMO technology typically uses only 2 or 4 antenna elements in an array, which is difficult to satisfy the ever-increasing communication requirement. Thus, massive MIMO was developed for the fifth-generation (5G) mobile communication systems. 
	For the upcoming 6G era, extremely large scale MIMO (XL-MIMO) \cite{ref:nearfield} is being investigated to achieve super spatial resolution and spectral efficiency. 
	However, simply increasing the number of antenna elements in an array may introduce new challenges to meet other key performance indicators (KPIs) for 6G, such as power and cost-efficiency. Developing innovative antenna architectures which enhanced performance in a cost-effective manner is necessary for 6G and beyond. Thus, 
	hybrid analog/digtial beamforming (HBF) \cite{ref:hbf}, lens antenna array \cite{ref:lens}
	sparse massive MIMO \cite{ref:spar}, fluid antenna \cite{ref:fluid} and movable antenna \cite{ref:mov} are investigated recently. 
	
	Recently, an alternative novel MIMO architecture termed ray antenna array (RAA) was proposed in \cite{RAA}. RAA is an innovative array architecture that arranges massive cheap antenna elements into multiple \emph{simple uniform linear arrays} (sULAs) with different orientations, each sULA corresponding to a ray. Unlike conventional ULA, all antenna elements within each sULA are directly connected without relying on any analog or digital beamforming. Thus, each sULA forms a beam towards the direction determined by its ray orientation. By dynamically choosing appropriate sULAs or rays, flexible beam can be achieved without relying on any analog \cite{ref:ana} or digital beamforming, i.e., no single phase shifter is needed, which is usually expensive especially in high-frequency systems like mmWave and TeraHertz. Compared with the classic HBF \cite{ref:hbf} architecture with ULA or \emph{uniform planar array} (UPA), which requires considerable phase shifters, a coarse estimation based on a sub-6 GHz communication system that employs the TGP2108-SM 6-bit digital phase shifter and the QM12002 RF switch shows that RAA only requires less than $1\%$ hardware cost \cite{RAA}. 
	
	\begin{figure}[!t]
		\centering
		\subfloat[Conventional cell sectoring system with 3 sectors. Users at sector boundaries typically have degraded performance.]{\label{con_sec}
			\includegraphics[width=0.8\linewidth]{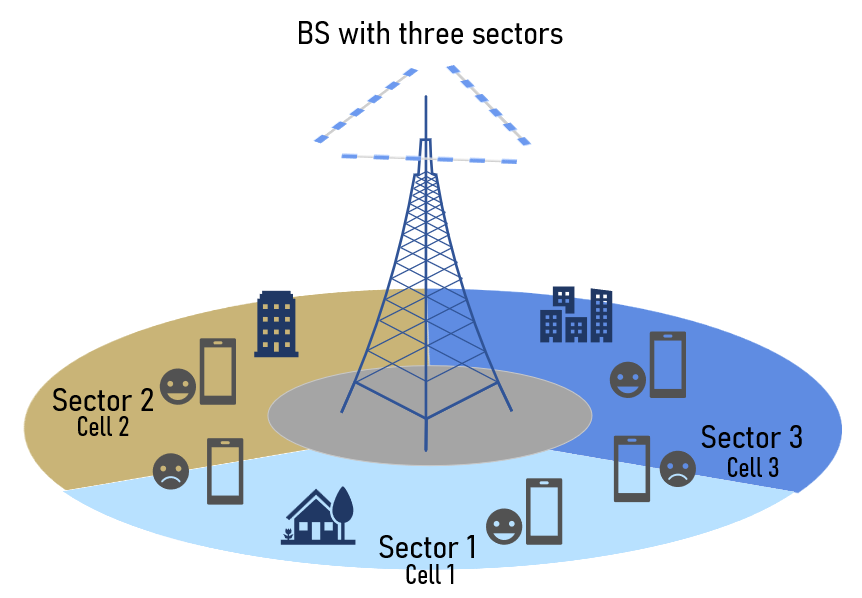}
		}
		\\
		\subfloat[Proposed omnicell wireless communication system with the full-angle RAA. Uniform angular resolution is achieved for all directions cost-effectively, and users for all directions are expected to achieve uniformly good services.]{\label{free_sec}
			\includegraphics[width=0.8\linewidth]{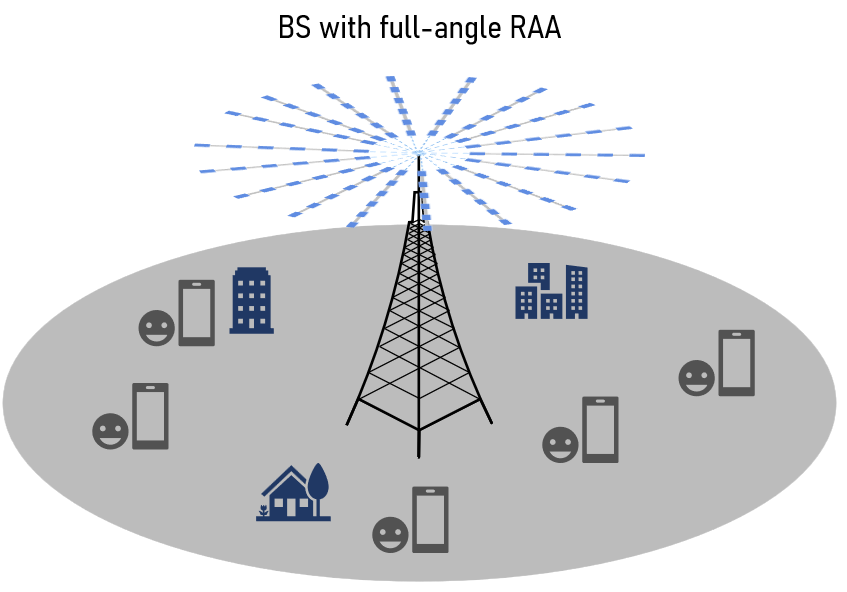}
		}
		\caption{Conventional cell sectoring system and proposed full-angle RAA enabled omnicell system. }
		\label{sector}
	\end{figure}	
	
	In this paper, to fully exploit RAA's promising advantages, we develop the concept of full-angle RAA, for which the sULAs span the whole angular range. 
	Furthermore, we propose a novel omnicell wireless communication paradigm by leveraging the full-angle RAA. This is significantly different from the conventional cell sectoring system, where		
	the spatial area is divided into multiple sectors, and each sector is served by an antenna array of the base station (BS) \cite{3GPP}. 
	For users close to the edge of a sector, as shown in Fig. \ref{sector}\subref{con_sec}, the performance usually degrades due to the beam overlapping of 
	2 adjacent sectors. 
	By contrast, for the proposed omnicell wireless communication system enabled by full-angle RAA, as illustrated in Fig. \ref{sector}\subref{free_sec}, uniformly high quality service can be achieved by deploying the BS at the center of a cell to fully take advantage of the full-angle RAA's uniform resolution for all directions. Compared with cell-free massive MIMO \cite{ref:cellfree}, the proposed omnicell wireless communication system does not require sophisticated coordination among different access points (AP). 
	
	To illustrate the performance of the proposed RAA-based omnicell wireless communication system, we compare its key performance indicators with conventional ULA/UCA-based cell sectoring system. 
	The results demonstrate that omnicell wireless communication has uniform angular resolution for all directions, which is significantly finer than the conventional cell sectoring system, especially for users close to the edge of conventional cell sectors. We derive the signal-to-interference-plus-noise ratio (SINR) of omnicell wireless communication system, and compare it with the ULA/UCA based cell sectoring. For multi-users communication, omnicell system achieves much higher achievable rate than the ULA/UCA-based cell sectoring system.
	
	\section{The Full-angle RAA}
\begin{figure}[!t]
		\centering
		\includegraphics[scale=0.6]{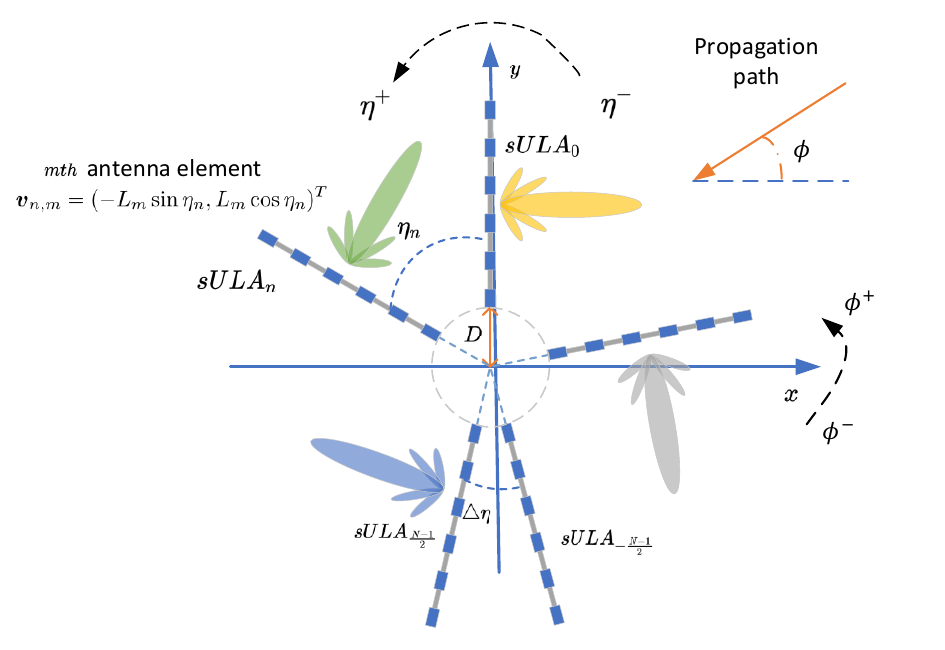}
		\caption{An illustration of full-angle RAA, where $N\times M$ antenna elements are arranged into $N$ rays, each having $M$ antenna elements. Each ray corresponds to one sULA with all antenna elements directly connected. Since each ray only needs to be responsible for a narrower angle, antenna element with higher directivity can be used compared to conventional antenna arrays.}
		\label{geo}
	\end{figure}

	In this section, we first present the RAA architecture proposed in \cite{RAA} and extend it to full-angle RAA that provides uniform angular resolution for $360^{\circ}$. As illustrated in Fig. \ref{geo}, a RAA is composed by $N\times M$ cheap antenna elements arranged into $N$ rays, each ray corresponding to one sULA that has $M$ elements separated by half wavelength. We use the term sULA to highlight the fact that different from the conventional ULA, all the $M$ elements within each sULA are directly connected without relying on any analog or digital beamforming. 
	The $nth$ ray has an orientation angle $\eta_n$ with respect to the positive y axis, where $n\in\mathcal{N}=\{-\frac{N-1}{2}, ..., 0, ...,\frac{N-1}{2}\}$. For the full-angle RAA, we have $-\pi \leq \eta_n \leq \pi$.
	
	The first antenna element of each ray has a distance $D$ to the origin so as to ensure that all antenna elements are separated by at least $\frac{\lambda}{2}$, where $\lambda$ is the wavelength. Thus, a full-angle RAA is characterized by the parameters $(N, M, D, \{\eta_n\}_{\mathcal{N}})$.
	For the $m{\rm th}$ antenna element in the $n{\rm th}$ ray, its location vector $\boldsymbol{v}_{n, m}$ is given by:
\begin{equation}
	\begin{aligned}
		\boldsymbol{v}_{n, m}=(-L_m\sin\eta_n, L_m\cos\eta_n)^T,
	\end{aligned}
\end{equation}
	where $L_m=D+(m-1)\frac{\lambda}{2}$. We consider a uniform plane wave (UPW) with direction vector denoted by $\mathbf k$. Note that different from that for the ray orientation, for notational convenience, we purposely choose the positive x axis as the reference direction for UPW. Therefore, for a UPW with angle of arrival (AoA) $\phi$, its direction vector is given by $\mathbf k=(\cos\phi, \sin\phi)^T$. The projection of $\boldsymbol{v}_{n, m}$ on $\mathbf k$ is given by:
	\begin{equation}
		\begin{aligned}
			\boldsymbol{v}_{n, m}^T\mathbf k&=-L_m\sin\eta_n\cos\phi+L_m\cos\eta_n\sin\phi\\
			&=D\sin(\phi-\eta_n)+\frac{\lambda}{2}(m-1)\sin(\phi-\eta_n).
		\end{aligned}
	\end{equation}

		Thus, for a UPW with AoA $\theta$, the array response vector of sULA with orientation $\eta_n$ is:
	\begin{align}
		\boldsymbol{a}(\theta, \eta_n)=\left(1,..., e^{j\pi(m-1)\sin\zeta_n}, ..., e^{j\pi(M-1)\sin\zeta_n}\right)^T,
		\label{a}
	\end{align}
	where $\zeta_n=\phi-\eta_n$ is the relative AoA with respect to the $nth$ ray. Therefore, the response matrix $\boldsymbol{A}(\phi)\in\mathbb{C}^{M\times N}$ is given by:
	\begin{align}
		\boldsymbol{A}(\phi)=\left[\boldsymbol{a}(\phi, \eta_n)\right]_{\mathcal{N}}\odot \left[b(\phi, \eta_n)\right]_{\mathcal{N}},
		\label{e}
	\end{align}	
	where $\odot$ represents Hadamard product and $b(\phi, \eta_n)$ denotes the phase shift of the first antenna element of each ray considering antenna element's radiation pattern, which is given by:
	\begin{align}
		b(\phi, \eta_n)=\sqrt{G(\zeta_n)}e^{j\frac{2\pi}{\lambda}D\sin(\zeta_n)},
		\label{b}
	\end{align}
	where $G(\zeta_n)$ refers to the radiation pattern of each antenna element.  
	It is observed from (\ref{a})-(\ref{b}) that the array response vectors of RAA depends on the relative AoA $\zeta_n$. Therefore, by selecting appropriate sULAs to be connected with the RF chains, a flexible beamforming towards the desired direction can be achieved without relying on conventional analog or digital beamforming technologies. 
	Since all the $M$ elements of each sULA are directly connected, the resulted output for all $N$ rays is denoted by $\boldsymbol{r}(\phi)=\boldsymbol{A}(\phi)^T \boldsymbol{1_{M\times 1}}$, given by:
	\begin{equation}
		\begin{aligned}
			\boldsymbol{r}(\phi)
			=\left[b(\phi, \eta_n)\right]_{\mathcal{N}}\odot \left[e^{j\frac{\pi}{2}(M-1)\sin\zeta_n}\frac{\sin(\frac{\pi}{2}M\sin\zeta_n)}{\sin(\frac{\pi}{2}\sin\zeta_n)}\right]_{\mathcal{N}}\label{dir}.
		\end{aligned}
	\end{equation}
	
	From (\ref{dir}), the null point of the $n{\rm th}$ sULA's beam lobe is determined by $\frac{\pi}{2}M\sin\zeta_n=k\pi$, where $k$ is an integer. To avoid adjacent sULAs from interfering with each other, the null point of the sULA's main lobe should be aligned to the peak point of the adjacent sULA's main lobe \cite{RAA}, so the orientation angle of the $n{\rm th}$ ray is set to:
	\begin{align}
		\eta_n=n\arcsin{\frac{2}{M}}\qquad n\in\mathcal{N}=\{0, ...,\pm\frac{N-1}{2}\}.
		\label{roa}
	\end{align}
	
	To separate the first antenna elements of the adjacent sULAs, the minimal distance between the first antenna element to the central point is given by:
	\begin{align}
		D\geq \frac{\lambda}{4\sin\left[0.5\arcsin(\frac{2}{M})\right]}.
		\label{Dreq}
	\end{align}	
	
	For the full-angle RAA to cover $360^{\circ}$, we have $\eta_{\frac{N-1}{2}}\approx \pi$. To ensure the distance between the first antenna elements of the $\frac{N-1}{2} {\rm th}$ and ${-\frac{N-1}{2}} {\rm th}$ sULA is larger than half wavelength, for any given $M$, the number of rays $N$ should satisfy:
	\begin{align}
		N=\left\lfloor
		\frac{2(\pi-\arcsin\frac{\lambda}{4D})}{\arcsin\frac{2}{M}}+1
		\right\rfloor.
		\label{nmcons}
	\end{align}	
	
	For $M\gg 1$, we have $D\approx\frac{M\lambda}{4}$, and $N\approx \lfloor M\pi\rfloor$. This approximation is validated to be effective when $M\geq4$.
	
	In summary, the characteristic parameters $(N, M, D, \{\eta_n\}_{\mathcal{N}})$ of the proposed full-angle RAA is uniquely identified by the following 3 steps.
	Firstly, determine the number of antenna elements $M$ in each sULA. Then, based on (\ref{Dreq}), determine the central distance $D$. Finally, determine the number of rays $N$ in the full-angle RAA based on \eqref{nmcons}.
	
	\section{proposed omnicell wireless communication System with Full-Angle RAA}
	In this section, we propose a novel system termed omnicell wireless communication system with the full-angle RAA. Based 
	on \eqref{dir}, the beamwidth of full-angle RAA is a constant 
	$2\arcsin\frac{2}{M}$ for all directions. Since sULA in the full-angle RAA only needs to be responsible for a narrow portion, we are able to select appropriate sULA to be connected with a RF chain based on the users' position without any beam lobe overlapping. 
	
	As shown in Fig. \ref{ula}, we compare the proposed omnicell system with the classic ULA-based cell sectoring system with three sectors, where each ULA has $M$ antenna elements and ULA serves a sector of angular range $\frac{2\pi}{3}$. The response vector of a ULA is given by:
	\begin{align} 
		\boldsymbol{a}_{ULA}(\phi)=\left(1, ..., e^{j\pi(m-1)\sin\phi}, ..., e^{j\pi(M-1)\sin\phi}\right)^T.
	\end{align}
	
	The DFT codebook of a ULA is given by:
	\begin{align}
		\boldsymbol{A}^{DFT}_{ULA}=[\boldsymbol{a}_{ULA}(\phi_n)]_{M\times N'},
	\end{align}
	where $N'$ is the number of DFT codeword and $\phi_n$ is the target direction angle of the $n{\rm th}$ codeword. The response pattern for the ULA using hybrid analog/digital beamforming (HBF) is given by:
	\begin{equation}
		\begin{aligned}
			\boldsymbol{r}_{ULA}(\phi)=\sqrt{G_{ULA}(\phi)}(\boldsymbol{A}^{DFT}_{ULA})^H\times \boldsymbol{a}_{ULA}(\phi)\\
			=\sqrt{G_{ULA}(\phi)}\left[\frac{\sin(\frac{\pi}{2}M(\sin\phi-\sin\phi_n))}{\sin(\frac{\pi}{2}(\sin\phi-\sin\phi_n))}\right]_{N'\times 1}.
			\label{hbf}
		\end{aligned} 
	\end{equation}
	
	For the omnicell wireless communication system with full-angle RAA, there are $N_{RF}$ RF chains to be connected with $N$ rays, a ray selection network (RSN) is developed in \cite{RAA} to find the selection matrix $\boldsymbol{S}\in\{0, 1\}^{N_{RF}\times N}$, which satisfies $\vert\vert [\boldsymbol{S}]_{i, :} \vert\vert^2 = 1$ and $\vert\vert [\boldsymbol{S}]_{:, n} \vert\vert^2 \leq 1$, $1\leq i \leq N_{RF}$ and $1\leq n \leq N$. We aim to maximize the spectral efficiency $R_{sum}$(bit/s/Hz) for $K$ users under different selection matrices, which is given by:
	\begin{align}
		R_{sum}=\sum_{k=1}^{K}\log_2(1+\text{SINR}_k).
		\label{ss}
	\end{align}	
	$\text{SINR}_k$ is the $k{\rm th}$ user's SINR. For Omnicell system, $\text{SINR}_k^{\text{RAA}}$ is given by:
	\begin{align}
		\text{SINR}_k^{\text{RAA}}=\frac{\bar{P_k}|\boldsymbol{f}_k^H\boldsymbol{S}\boldsymbol{h}_k|^2}{\bar{P_t}\sum_{i\neq k}|\boldsymbol{f}_k^H\boldsymbol{S}\boldsymbol{h}_i|^2+M\vert\vert\boldsymbol{f}_k^H \boldsymbol{S}\vert\vert^2},
		\label{sinrraa}
	\end{align}
	where $\boldsymbol{h}_k$ is the channel vector of the $k{\rm th}$ user who has $L_k$ paths, which is given by:
	\begin{align}
		\boldsymbol{h}_k=\sum_{l=1}^{L_k}\alpha_{k, l}\boldsymbol{r}(\phi_{k, l}),
		\label{h}
	\end{align}
	$\alpha_{k, l}$ is the complex gain factor of the $k{\rm th}$ user's $l{\rm th}$ path. $\bar{P_k}=\frac{P_k}{\sigma^2}$ is the transmit signal-to-noise ratio (SNR) and $P_k$ is the transmit signal power of the $k{\rm th}$ user. The circularly symmetric Gaussian noise vector of the full-angle RAA is given by $\boldsymbol{z}\sim N_{\mathbb{C}}(0, M\sigma^2\boldsymbol{I}_{N})$, so the received noise power is given by $M\sigma^2\vert\vert\boldsymbol{f}_k^H \boldsymbol{S}\vert\vert^2$, where $\boldsymbol{f}_k\in\mathbb{C}^{N_{\text{RF}}}$ is the baseband beamforming vector for the $k{\rm th}$ user. The optimization problem to find the selection matrix is given by:
	\begin{equation}
		\begin{aligned}
			\max_{\boldsymbol{F, S}} \quad & R_{sum}=\sum_{k=1}^{K}\log_2(1+\text{SINR}_k)\\ 
			{\rm s.t.} \quad & (C1): \boldsymbol{S}\in\{0, 1\}^{N_{RF}\times N}\\
			& (C2): \vert\vert \boldsymbol{S}_{i, :}\vert\vert^2=1, 1\leq i \leq N_{RF}\\
			& (C3): \vert\vert \boldsymbol{S}_{:, j}\vert\vert^2\leq1, 1\leq j \leq N\\
			& (C4): \boldsymbol{F}=(\boldsymbol{f}_1,\boldsymbol{f}_2,\cdots,\boldsymbol{f}_K),\\ 
			\label{opt}
		\end{aligned}
	\end{equation}
	where $\boldsymbol{F}$ is the baseband beamforming matrix for all $K$ users. To find a selection matrix for (\ref{ss}), an efficient greedy algorithm is proposed in \cite{RAA}.
	By using directional antenna elements, i.e., the radiation pattern satisfies 3GPP directional antenna model \cite{3GPP}:
\begin{subequations}\label{antenna}
	\begin{equation}
		\begin{aligned}
			G(\zeta_m)=
			-\min\left[-A_{dB}\left(\zeta_m\right), A_{max}\right]
		\end{aligned}
	\end{equation}
	
	\begin{equation}
		\begin{cases}
			&A_{dB}\left(\zeta_m\right)=-\min\left[12\left(\frac{\zeta_m}{\zeta_{3dB}}\right)^2,A_{max}\right]\\[8pt]
			
			&A_{max}=30\text{dB}\\[8pt]
			&\zeta_{3dB}=65^{\circ}\\[8pt]
		\end{cases}
	\end{equation}
\end{subequations}
	From (\ref{dir}) and (\ref{hbf}), full-angle RAA's response pattern includes AoA $\zeta_n=\phi-\eta_n$ in a $\sin(\cdot)$ function, so the full-angle RAA will sample the angle domain by angle deviation $\arcsin(\frac{2}{M})$ \cite{RAA}, while the ULA-based cell sectoring using HBF samples the sine value of the angle domain, i.e., $\sin\phi$, by $\frac{2}{M}$. Because of the nonlinear characteristics of $\sin(\cdot)$ function, ULA-based cell sectoring HBF's beamwidth in large angle is wider compared with full-angle RAA, and its power gain is lower than full-angle RAA since the directivity of antenna elements.
	
	In the following part, we will compare another UCA-based cell sectoring technology and the omnicell wireless communication system. Similar to the full-angle RAA, UCA can also achieve uniform resolution for all directions \cite{ref:semi_array} by using beamforming technology. 
	We also divide UCA-based cell sectoring system's serve region into three $\frac{2\pi}{3}$ sectors.  	
	To make a fair comparison, we choose a UCA that has $N$ antenna elements and, each antenna element's orientation angle is given by:
	\begin{align}
		\phi_n=\frac{2\pi}{N}(n-1)\qquad (n=1, 2, ..., N).
	\end{align}
	The radius $a$ of the UCA satisfies \cite{ref:semi_array} :
	\begin{align}
		a=\frac{N\lambda}{4\pi}.
		\label{rad}
	\end{align}
	Thus, the response vector of UCA is given by:
	\begin{align}
		\boldsymbol{a}_{UCA}(\phi)=\left(
		1, ..., e^{j\frac{4\pi}{\lambda}a\sin\left(
			\frac{2\phi-(\phi_1+\phi_n)}{2}
			\right)\sin\left(
			\frac{\phi_n-\phi_1}{2}
			\right)}, ...
		\right)^T.
		\label{uca}
	\end{align}
	
	From (\ref{uca}), the elements' phase change in a nonlinear manner, so it is impossible to use linear codebook like DFT codebook of ULA to perform beamforming to a UCA. We may use the codebook termed parametric codebook \cite{ref:parametric,ref:signal} to compare its performance with the full-angle RAA. That is, the angle that corresponds to the $n{\rm th}$ codeword $\phi_n$ is given by $\phi_{n+1}=\phi_n+\chi$, 
	where $\chi$ is a variable parameter. To make a fair comparison, we set $\chi=\arcsin(\frac{2}{M})$, so the codebook's angle deviation is the same as the proposed full-angle RAA. The parametric codebook is given by:
	\begin{align}
		\boldsymbol{A}_{UCA}=[\boldsymbol{a}_{UCA}(\phi_n)]_{\mathcal{N}}\qquad (n=1,2,\cdots,N).
	\end{align}
	The response pattern of the UCA $\boldsymbol{r}_{UCA}(\phi)\in \mathbb{C}^{N}$ is given by:
	\begin{align}
		\boldsymbol{r}_{UCA}(\phi)=\sqrt{G_{UCA}(\phi)}\boldsymbol{A}^H_{UCA}\times \boldsymbol{a}_{UCA}(\phi).
		\label{ucares}
	\end{align}
	Using isotropic antenna elements in the UCA, $\sqrt{G_{UCA}(\phi)}=1$. To reach such an all directions beamforming, UCA will select  $\frac{N-1}{2}$ or $\frac{N+1}{2}$ antenna elements in the array facing to the codeword's target angle $\phi_n$, while the proposed full-angle RAA can use the directional antenna elements to focus energy to one direction since each sULA only need to be responsible for a narrow portion. So it is clear that though UCA can generate beam lobes whose shape is similar to full-angle RAA, it can not reach similar power gain as the full-angle RAA with the same number of rays as the antenna array elements. UCA-based cell sectoring costs more than ULA-based cell sectoring since every antenna element in the UCA needs a phase shifter.
	
	In the following part, we will compare the maximal SINR and sum rate performance of omnicell wireless communication system with full-angle RAA and ULA/UCA-based cell sectoring system. 
	We have already developed the SINR of the omnicell wireless communication with full-angle RAA in section \Rmnum{3} by \eqref{sinrraa}. 
	
	Similar to RAA, we introduce a selection matrix $\boldsymbol{S}'$ to represent the beamforming codeword selection process. Therefore, the $k {\rm th}$ user's SINR is given by: 
	\begin{align}
		\text{SINR}_k=\frac{\bar{P_t}|\boldsymbol{f}^H_k\boldsymbol{S}'(\boldsymbol{A}^H\boldsymbol{h}_k^{p_k})|^2}{\bar{P_t}\sum_{i\neq k}|\boldsymbol{f}_k^H\boldsymbol{S}'(\boldsymbol{A}^H\boldsymbol{h}_i^{p_k})|^2+M\vert\vert\boldsymbol{f}_k^H\boldsymbol{S}' \vert\vert^2},
	\end{align}		
	where $\boldsymbol{A}$ is the DFT codebook for ULA-based cell sectoring or parametric codebook for UCA-based cell sectoring and $\boldsymbol{h}_k^{p_k}$ is the corresponding channel vector for the ULA/UCA covering the $p_k$ sector, which is defined by $\boldsymbol{h}_k^{ULA, p_k}=\sum_{l=1}^{L_k}\alpha_{k, l}\boldsymbol{a}^{ULA}(\phi_{k, l}^{p_k})$ or $\boldsymbol{h}_k^{UCA, p_k}=\sum_{l=1}^{L_k}\alpha_{k, l}\boldsymbol{a}^{UCA}(\phi_{k, l}^{p_k})$. We aim to find the selection matrix $\boldsymbol{S}'$ and $K$ users' beamforming matrix $\boldsymbol{F}=[\boldsymbol{f}_k]\in \mathbb{R}^{N_{RF}\times K}$ which meets highest sum rate similar to RAA in \eqref{opt} by exhaustive search or greedy algorithm, which is validated to be effective in \cite{RAA}.
	
	\begin{figure}[!t]
		\centering
		\includegraphics[scale=0.44]{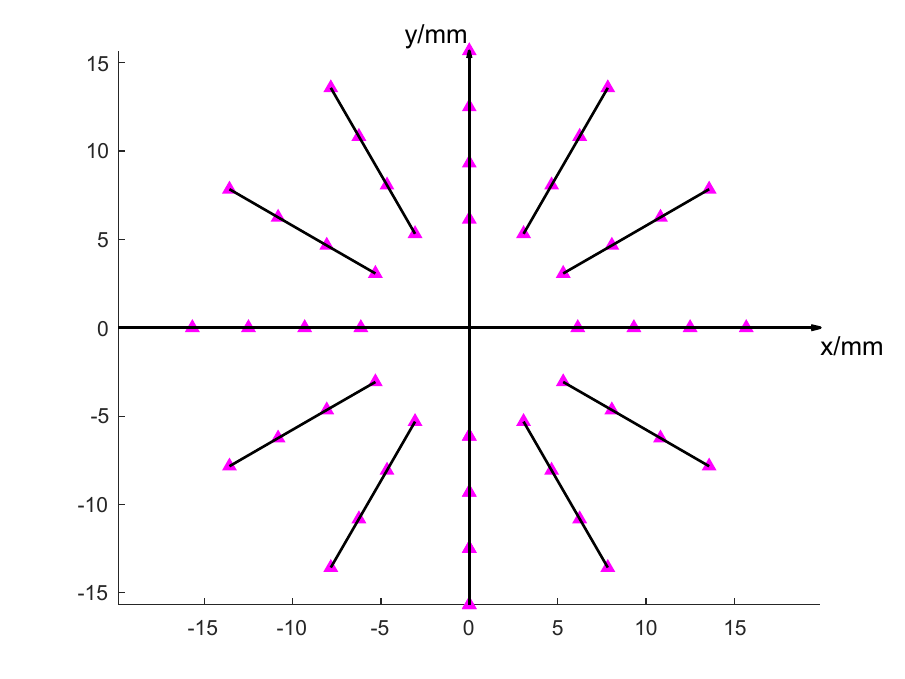}
		\caption{An illustration of full-angle RAA at $f_c={47.2}\text{GHz}$ with $M=4$, $N=13$, $D=6.1\text{mm}$ and $\eta_n=n\arcsin(\frac{1}{2})$}
		\label{raa}
	\end{figure}
	
	\section{Simulation results}
	In this section, we will compare the resposne pattern, SINR performance and cost-efficiency of the omnicell system and the cell sectoring system.
	\subsection{Response Pattern of the Omnicell System}
\begin{figure}[htbp]
		\centering
		\includegraphics[scale=0.5]{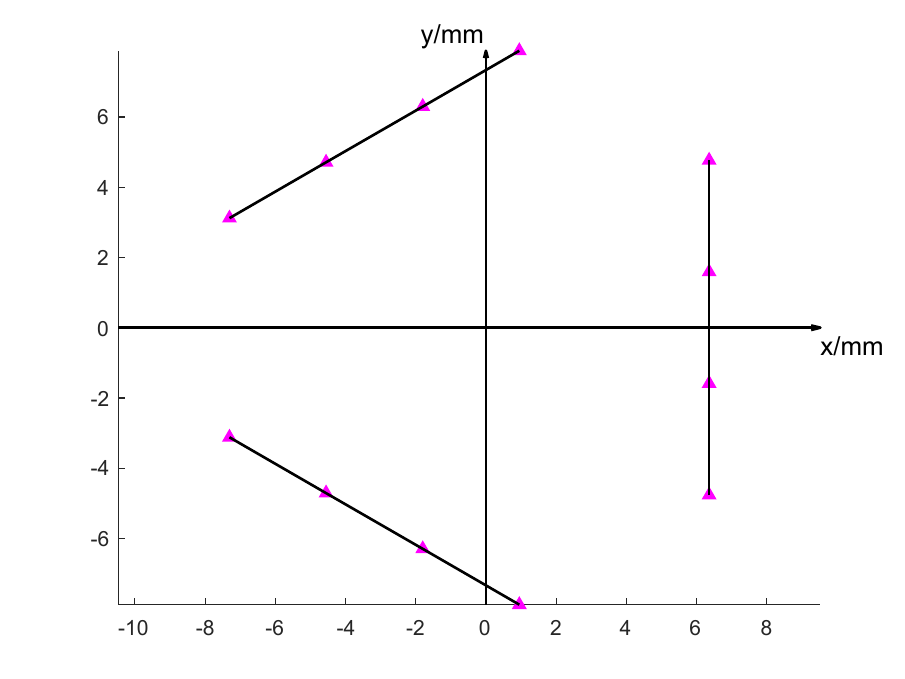}
		\caption{ULA-based cell sectoring array with 3 ULAs using HBF, each ULA has 4 elements and serves a sector of $\frac{2\pi}{3}$}
		\label{ula}
	\end{figure}

	To illustrate the beam pattern of omnicell system and the UCA/ULA-based cell sectoring clearly, we consider an omnicell system enabled by a full-angle RAA with few antenna elements $M=4$, $N=13$ working under central frequency $f_c={47.2}\text{GHz}$. From (\ref{Dreq}), $D={6.1}\text{mm}$ and $\eta_n=n\frac{\pi}{6}$, as illustrated in Fig. \ref{raa}, here we set the intersection of 3 sectors as the origin point. 
	To compare the response pattern of the full-angle RAA with the ULA-based cell sectoring as the benchmark, we set a ULA-based cell sectoring with 3 sULAs and each of them serves a sector of $\frac{2\pi}{3}$, as shown in Fig. \ref{ula}.
	
\begin{figure}[!t]
		\includegraphics[width=\linewidth]{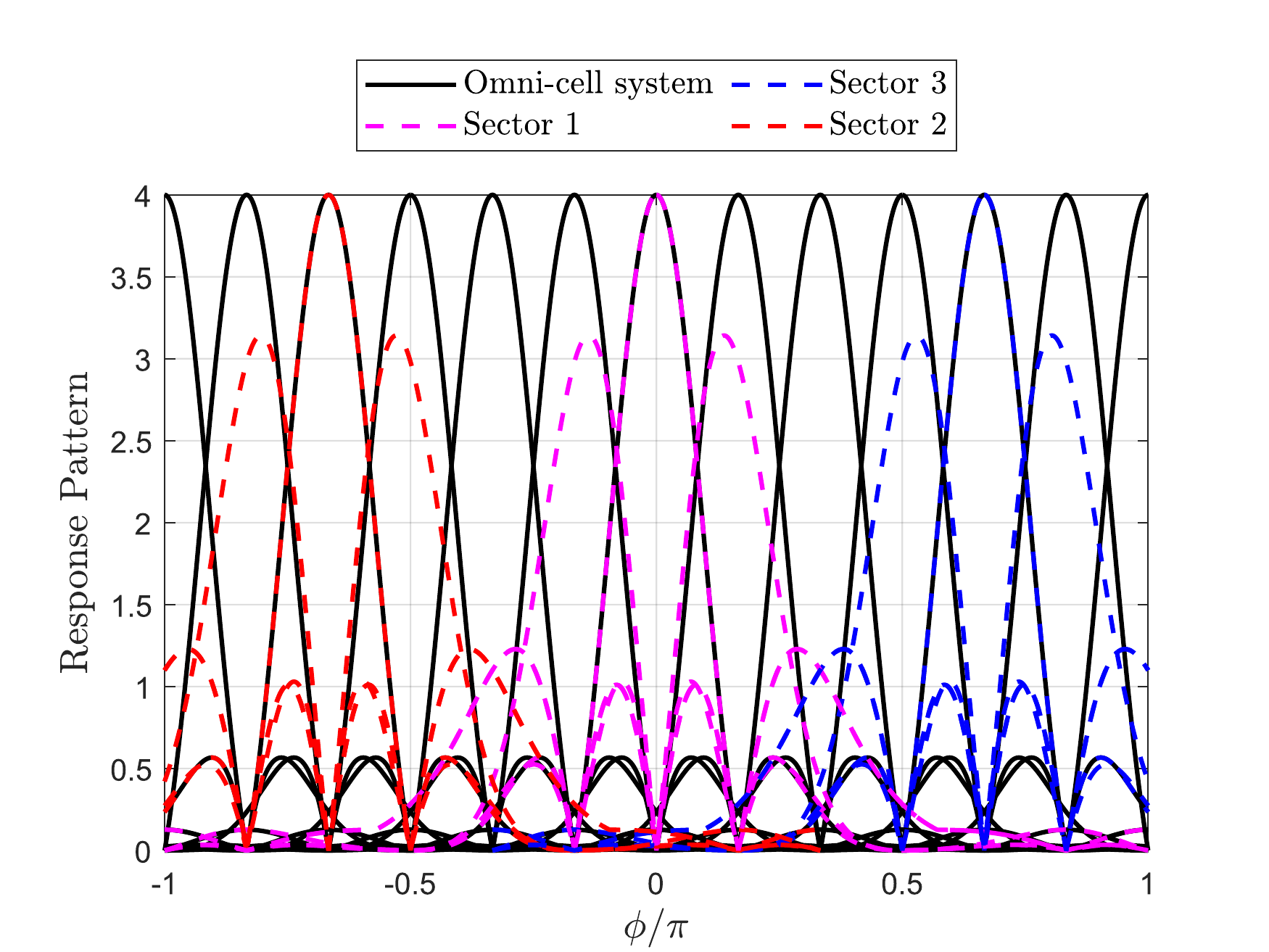}
		\caption{Response pattern of the omnicell wireless communication system enabled by a full-angle RAA with $(N, M)=(13, 4)$ and the ULA-based cell sectoring with AoA $\phi$ varies from $-\pi$ to $\pi$}
		\label{respat}
	\end{figure}

	The response patterns $|r_{\eta_n}(\phi)|$ and $|r_{ULA, \phi_{n'}}(\phi)|$ in \eqref{dir} and \eqref{hbf} is illustrated in Fig. \ref{respat}. It is obvious that for the wave coming around $|\phi|=\frac{\pi}{3}$ or $|\phi|=\pi$, the omnicell system's beam lobe is higher than conventional ULA-based cell sectoring, and latter one's beam lobe overlaps with adjacent lobes.
	Based on the generally used HBF codeword selection algorithm, the ULA-based cell sectoring system's inter-sector interference is very high due to the low beam amplitude and overlapping, thus, intra-sector and inter-sector interference is suppressed with omnicell system.
	
\begin{figure}[!t]
		\centering
		\includegraphics[width=\linewidth]{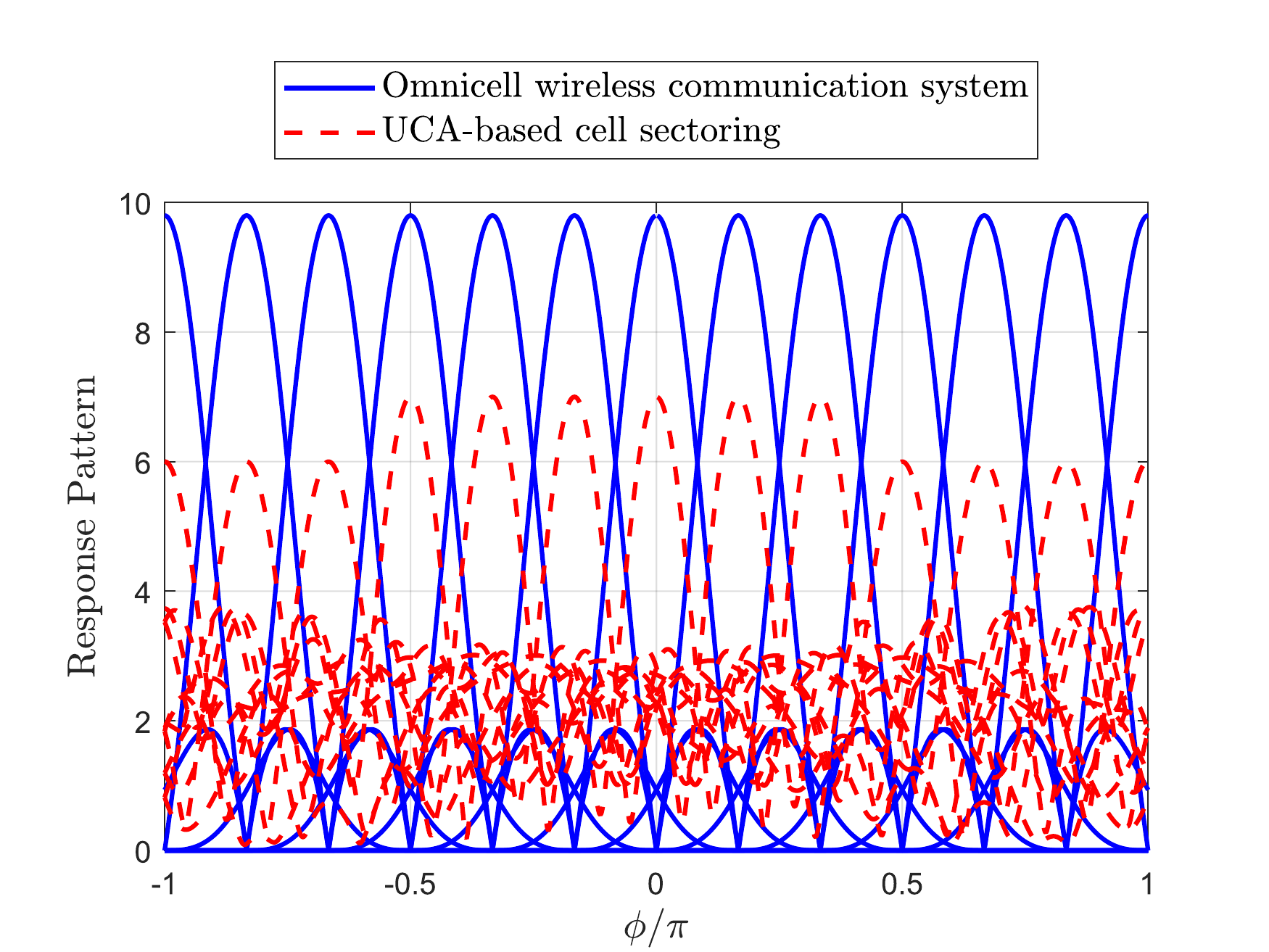}
		\caption{Response pattern of the omnicell system and UCA-based cell sectoring system with $(N, M)=(13, 4)$ and UCA with AoA $\phi$ varies from $-\pi$ to $\pi$}
		\label{pat2}
	\end{figure}

	Furthermore, we consider a UCA-based cell sectoring with $N=13$ antenna elements uniformly distributed in a 
	circular array. From \eqref{rad}, the radius of the circle is $a=D={6.1}\text{mm}$. 
	The response patterns $|r_{UCA, \phi_{n'}}(\phi)|$ and $|r_{\eta_n}(\phi)|$ in \eqref{ucares} and \eqref{dir} are illustrated in Fig. \ref{pat2}. By using directional antenna elements in the full-angle RAA, the power gain of the full-angle RAA is higher than that of UCA. 
	
	\subsection{SINR and Sum Rate Performance of Omnicell System}
	
	\begin{table}[htbp]
		\centering
		\caption{Simulation parameters of 3GPP Uma NLOS scenario \cite{3GPP}}
		\label{tab:sim}
		\begin{tabular}{|c|c|c|}
			\hline
			\textbf{Parameter} & \textbf{Symbol} & \textbf{Definition / Value} \\
			\hline
			Central frequency & $f_c$ & $47.2\text{GHz}$\\ 
			\hline
			Number of users & $K$ & 10\\
			\hline
			Number of RF chains & $N_{\text{RF}}$ & 10\\
			\hline
			Number of clusters & $N_c$ & 20\\
			\hline
			Number of rays per cluster & $N_r$ & 20\\
			\hline
			\makecell{Number of paths \\of the $k\text{th}$ user} & $L_k$ & \makecell{$L_k=N_c\times N_r=400$}\\
			\hline
			\makecell{LOS azimuth angle\\ of the $k\text{th}$ user} & $\theta_k$ & \makecell{$\theta_k\sim Uniform(-\pi,\pi)$}\\
			\hline
			\makecell{Transmit SNR\\ of the $k\text{th}$ user} & $(\bar{P_t})_k$ & \makecell{$-10\text{dB}\leq\bar{P_t}\leq10\text{dB}$\\ 
				($\bar{P_t})_k=\bar{P_t}$}\\
			\hline
			\multicolumn{3}{|c|}{
			\makecell{For paths AoA parameters relative to angle spread and
			complex\\ power gain factors relative to cluster delay, please refer to \cite{3GPP,RAA_long}.}
			}\\
			\hline
		\end{tabular}
	\end{table}
	
\begin{figure}[!t]
	\centering
	\includegraphics[width=\linewidth]{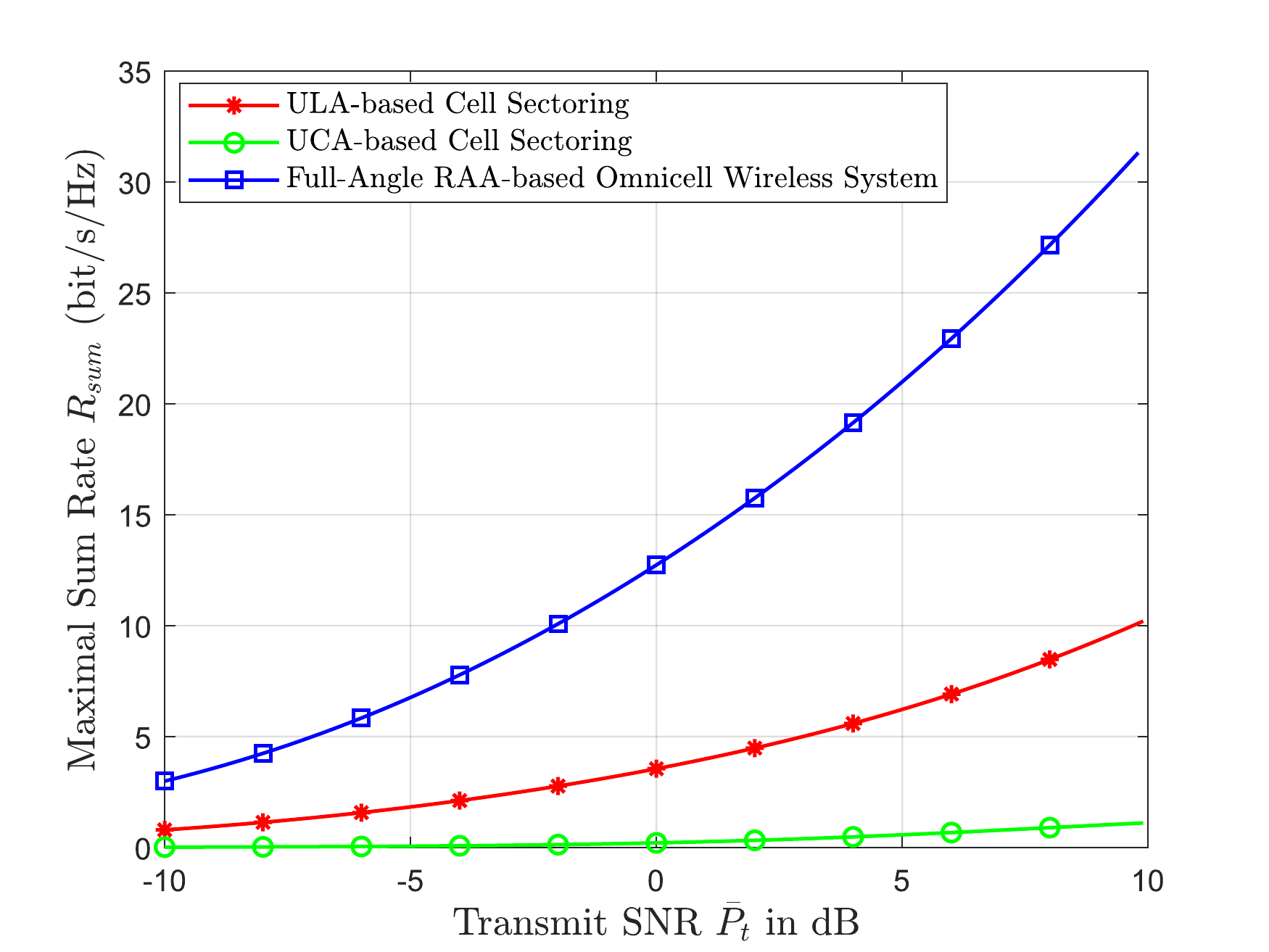}
	\caption{10-users maximal sum rate for the omnicell wireless communication system, UCA/ULA-based cell sectoring. Proposed omnicell wireless communication system has larger sum rate than ULA/UCA-based cell sectoring with the same transmit SNR.}
	\label{snrmul}
	\end{figure}

	Finally, we will compare the SINR performance of the omnicell wireless communication system and the 
	conventional cell sectoring system. The simulation parameters are set to satisfy 3GPP Uma NLOS scenario \cite{3GPP} and are given in Table \ref{tab:sim}. 
	The omnicell system uses a full-angle RAA with $M = 64$ and $N = 201$; the ULA-based cell sectoring system equips 64 elements in a ULA and the UCA-based cell-sectoring system has 100 antenna elements in the array. 	Using minimum mean square error (MMSE) equalization baseband beamforming, i.e.,  $\boldsymbol{f}_k=\boldsymbol{C}_k^{-1}\boldsymbol{S}\boldsymbol{h}_k$ with $\boldsymbol{C}_k = \boldsymbol{S}(\sum_{i\neq k}\boldsymbol{h}_i\boldsymbol{h}_i^H+\frac{M}{\bar{P_t}}\boldsymbol{I}_N)\boldsymbol{S}^T$, the maximal sum rate $R_{sum}$ of omnicell wireless communication system and UCA/ULA-based cell sectoring is illustrated in Fig. \ref{snrmul}. It shows that omnicell system has larger sum rate than UCA/ULA-based cell sectoring, which is attributed to the low intra-sector inter-user interference brought by the directional power gain of each antenna element since we design the optimized selection matrix to find the sULA which is determined by optimization problem \eqref{opt}. 
	It is further proved that the greedy algorithm designed for optimizing selection matrix $\boldsymbol{S}$ can be reduced to a simple minimum angle distance problem \cite{RAA_long}, thus decreasing computing resources cost and power consumption. 
		
	\subsection{Cost-efficiency Analysis}
	For the architecture discussed in the previous section, 
	we assume a 5G mmWave communication systems that employs the ADMV4728 beamformer and the TGS4302 RF switch. From manufacture's quotation, one phase shifter costs more than ${120}\text{\$} $ \cite{ref:shifter}, one RF switch costs ${28.62}\text{\$}$ \cite{ref:switch} and one cheap antenna element costs no more than ${0.01}\text{\$}$. The total cost of the omnicell system with full-angle RAA is $\text{cost}_{RAA}=NM\text{cost}_{\mathrm{antenna}}+N_{RF}N\text{cost}_{\mathrm{switch}}/2$ and the cost of sector covering ULA is $\text{cost}_{ULA}=N_{RF}M\text{cost}_{\mathrm{shifter}}+3M\text{cost}_{\mathrm{antenna}}$. Omnicell system costs $\text{cost}_{RAA}={28,892}\text{\$}$ and ULA-based cell sectoring costs $\text{cos}_{ULA}={76,801}{\$}$; thus, we achieve better performance as well as decrease hardware cost to $37.62\%$. Sensitivity analysis shows that the omnicell system is more cost-efficient for antenna element whose price is no more than 3.8\$.
	
	\section{conclusion}
	In this paper, we propose a novel antenna architecture termed full-angle RAA and an innovative omnicell system with the full-angle RAA. Without relying on conventional beamforming technology, the omnicell system costs much less than UCA/ULA-based cell sectoring. And the omnicell system can achieve better spatial resolution than cell sectoring system. Owing to low interference with the same transmit SNR, omnicell system has larger sum rate than conventional UCA/ULA-based cell sectoring. The highlights of the omnicell system mean that it is a feasible solution for unmanned aerial vehicle (UAV) swarm communication and environment sensing which need to realize high spatial resolution and large sum rate in low transmit SNR with the lowest possible cost.

\bibliographystyle{IEEEtran}
\bibliography{IEEEabrv,myrefs}

\end{document}